\documentclass[journal,transmag]{IEEEtran}
\usepackage{graphicx}
\usepackage{graphics}
\usepackage{subfigure}
\usepackage{calligra}

\begin{document}
%
\title{Spacetime as the optimal generative network of quantum states: a roadmap to QM=GR?}



%

\author{\IEEEauthorblockN{Xiao Dong, Ling Zhou}
\IEEEauthorblockA{Faculty of Computer Science and Engineering, Southeast University, Nanjing, China}}

%



\IEEEtitleabstractindextext{%
\begin{abstract}
The idea that spacetime geometry is built from quantum entanglement has been widely accepted in the last years. But how exactly the geometry is determined by quantum states is still unclear. In this note based on the idea of deep learning, we propose a mechanism for Susskind's QM=GR hypothesis, \emph{spacetime geometry as the optimal generative network of quantum states}. We speculate that the space geometry stems as a geodesic tensor network which defines the quantum state complexity of a fundamental quantum state under a given metric. Spacetime corresponds to an evolving tensor network that generates a time evolutional fundamental quantum system. This mechanism provides (a) a constructive correspondence between quantum states and spacetime geometry; (b)a spacetime structure emerging from a highly constrained geodesic so that the QEC-like structure shown in AdS/CFT can be naturally realized and (c) a mechanism to derive the gravity equation from the concept of quantum state complexity. With this mechanism, spacetime can have a quantum mechanical description. We hope this may lead to another view direction to understand the basic rules of our world. \end{abstract}

}

\maketitle



\IEEEdisplaynontitleabstractindextext

%
\IEEEpeerreviewmaketitle

\section{Motivation}

With the recent development of quantum information, more and more physicists tend to accept the concept that spacetime emerges from quantum information. Ideas on this track include ER=EPR, complexity=volume, complexity=action, RT formulae, holographic error correcting code, exact holographic mapping(EHM), AdS/MERA duality, bulk entanglement gravity, computational universe etc. Recently Susskind claimed QM=GR, which aims to build a correspondence between \emph{any} quantum mechanical system with a gravitational system. All the above works have the same belief that (a)Spacetime geometry origins from quantum states and (b) Spcetime emerges from quantum states through a quantum computation or a tensor network determined by quantum states. Since quantum circuits can be represented by tensor networks, in this paper we will only use tensor networks.

To understand how a spacetime geometry is built from quantum states (and their evolution), we have now at least two testing benches. The first is the famous AdS/CFT duality, which shows a correspondence between a boundary CFT and a gravitational bulk system. Another system is the bulk gravitation from Hilbert space without a boundary or Bulk Entanglement Gravity(BEG)\cite{Cao2017Bulk}, which aims to directly derive a gravitation compatible spacetime from \emph{fundamental} quantum states in a Hilbert space. Obviously the fundamental law of building spacetime geometry should build the spacetime from the background-free QM system, which we call the \emph{fundamental quantum states}. The AdS/CFT can then be regarded as either a component of the QM=GR or a deduced result of the QM=GR mechanism. In this work we would consider them both. Our idea is that the relatively simpler but well-studied AdS/CFT can help to \emph{guess} the mechanism of QM=GR. On the other hand, it can also work as a verification of any QM=GR mechanism if our assumption that the AdS/CFT should be deduced from the QM=GR is correct. This work flow remind us of the generative adversarial network (GAN)\cite{Goodfellow2014Generative} structure in deep learning, where the QM=GR is the generative model and the AdS/CFT plays the role of both the training data and part of the discriminative model.

QM=GR as the generative model of emergent spacetime, its input domain is a background free Hilbert space and the output includes a spacetime geometry, a matter field and a gravitational equation. Inspired by the observations that spacetime emerges from quantum information through a tensor network, the generative model QM=GR consists of two components: a mapping from quantum states to tensor networks and a mapping from tensor networks to gravity and spacetime geometry. To be more specific, if we first only consider space geometry, given a quantum state $\rho$, its correspondent space geometry emerges by a two-step mapping as $\rho\rightarrow TN(\rho)\rightarrow Geo(TN(\rho))$, where $TN(\rho)$ and $Geo(\rho)$ are the tensor network and the the space geometry determined by $\rho$. If spacetime geometry is considered, then the input is a curve in the Hilbert space $\rho(s)$, and the spacetime is generated by $\rho(s)\rightarrow NT(\rho(s))\rightarrow Geo(NT(\rho(s)))$. An ideal QM=GR mechanism should at least achieve the following goals
\begin{itemize}
  \item \textbf{Task1}: A mapping between a quantum state and a tensor network;
  \item \textbf{Task2}: A procedure to build a spacetime geometry from the varying tensor network when the fundamental quantum state evolves;
  \item \textbf{Task3}: The emergent spacetime satisfies gravitational equation;
  \item \textbf{Task4}: Support the AdS/CFT duality.
\end{itemize}

In this paper we will scratch a roadmap for the QM=GR mechanism and give a proof-of-concept of it. Though the details of the mechanism is not very clear, we have already found lots of mosaics, from which a clearer picture begins to emerge.

\begin{itemize}
  \item \textbf{BEG}: BEG \cite{Cao2017Bulk}\cite{Cao2017Space} aims to directly build a geometry from a redundancy constrained (RC) fundamental quantum state without the intermediate tensor network. The spatial geometry is defined by associating areas with the entanglement entropy between subsystems and the evolution of the RC state traces out a spacetime geometry. Similar to AdS/CFT, a generalized entanglement equilibrium is derived from a generalized Ryu-Takayanagi (RT) formula and the entanglement first law(EFL), which shows the emergent spacetime obeys a linearized Einstein's equation. But in BEG, how to construct the spacetime geometry explicitly from the fundamental state is still missing. Also it only works on RC states and therefore is not a complete QM=GR model.

  \item \textbf{AdS/CFT}: In AdS/CFT, different tensor networks were proposed to understand the emergent bulk system including MERA/AdS duality\cite{Swingle2009Entanglement}\cite{Swingle2012Constructing}, holographic quantum error-correcting code(HQEC)\cite{Pastawski2015Holographic}, bidirectonal holographic codes(BHC)\cite{Yang2016Bidirectional}, random tensor networks(RTN)\cite{Hayden2016Holographic}, exact holographic mapping (EHM)\cite{Qi2013Exact} etc. Their common idea is to design special erasure QEC-like tensor networks to support RT relations and the wedge/entanglement wedge reconstruction of bulk operators. Also gravitational equation can emerge from the RT formula, EFL and entanglement equilibrium. The main problem is the tensor works used here are hand-crafted except the MERA/AdS duality. So an emergent tensor network, which shows a erasure code property, from the boundary CFT state is desired.

  \item \textbf{Surface/state correspondence}: The surface/state duality\cite{Miyaji2015Surface} is a proposal to link the bulk geometry encoded in a tensor network with quantum states, which is applicable to both QM=GR and AdS/CFT. What's more, the local bulk geometry can be derived by information metric due to the surface/state correspondence.

  \item \textbf{Quantum complexity}: Quantum complexity is playing a more and more important role in the spacetime structure. Susskind's complexity=volume and complexity=action hypothesises have been used to derive Einstein's equation from the variation of quantum complexity\cite{Czech2018Einstein}\cite{Ge2018Quantum}\cite{Matsueda2014Derivation}. In \cite{Dong_time} quantum state complexity was regarded as the origin of the thermodynamic arrow of time.  These observations give a strong indication that QM=GR is closely related with quantum complexity, which is exactly our idea.

  \item \textbf{Computational universe}: In \cite{Lloyd2006A}\cite{Lloyd2012A} Lloyd proposed to build spacetime geometry from a quantum computation, which is consistent with Einstein's equation. This is an evidence that quantum computation/tensor network can be used to build spacetime geometry, which is a key component of QM=GR.

  \item \textbf{Quantum information and gravity}: In \cite{Matsueda2013Emergent}\cite{Matsueda2014Derivation} it's found that the second derivative of the entanglement entropy directly represents the spacetime metric in AdS/CFT, so that AdS/CFT can be understood as to storage quantum entanglement in classical spacetime. In other words, the bulk geometry encodes the boundary quantum state.

  \item \textbf{Tensor network and geometry}: Loop quantum gravity (LQP) aims to build background-independent spacetime geometry from spin network and spin foam. Recently \cite{Han2017Loop} shows that tensor networks are emergent from spin network by coarse graining. Therefore in AdS/CFT both the RT formula and the emergent bulk geometry can be derived from spin networks. The introduction of spin network provides a passway to build spacetime geometry from tensor networks. Another work relating tensor network with geometry \cite{Evenbly2011Tensor} addressed the geometry of ground states of local Hamiltonians on a D dimensional lattice, where the relationship between the holographic geometry built on the tensor network to represent the ground state and the physical geometry generated by the pattern of the interaction of the Hamiltonian was discussed.

  \item \textbf{Deep learning}: Deep learning seems not directly related with emergent spacetime. But we believe the idea of deep learning might be essential to understand QM=GR. In fact deep learning and neural networks have been used for efficient representation of many-body quantum states, spatial geometry learning from entanglement entropy/mutual information of quantum states. Also our former work showed a similarity between the geometrical structure of quantum computation and deep learning systems\cite{Dong_deep}. In QM=GR, we are interested in the generative property of deep networks, which means a deep network with a certain \emph{structure} can be regarded as a generator of a physical system. For example in image processing, low level image processing tasks can be accomplished by an untrained deep network\cite{Ulyanov2017Deep}. Generative adversarial network(GAN) also has been used to generate images of human faces and other objects\cite{Karras2017Progressive}. This means the space of natural images are encoded in the structure of a deep network, and therefore we have a correspondence between the configuration of a physical system and a generative model. The generative network works as both a generator and a projection operator to the state space of the correspondent physical system. This is very similar to the idea that the structure of a tensor network encodes the space of certain quantum states such as the ground state of local Hamiltonian. From this point of view, tensor networks are both generators of quantum states and spacetime geometry. Or as in \cite{Ulyanov2017Deep} spacetime is the structure of the generative model of the quantum entanglement of quantum states. Therefore quantum states and spacetime geometry are just the two sides of the same coin and QM=GR now has a concrete foundation. Another observation from the deep learning field is, different images can be generated either by a variable input on a fixed deep network or a fixed input on a deep network with a fixed structure but variable parameters. As an analogue, if a quantum state is generated by a tensor network, the bulk freedoms include both the structure and the parameters of the network. This is slightly different from the current toy models on AdS/CFT, where the tensor network structure seems to be fixed and the so called bulk freedoms are only the dangling bulk legs. We believe the dynamics of the bulk system involves both the structure and the parameters of the tensor network just as in general relativity, where the dynamics includes both geometry and matter.
\end{itemize}

Obviously the key component of QM=GR is a tensor network, which on one hand is a representation/generator of the fundamental quantum state and on the other hand it generates the geometry. The tensor network correspondent to a fundamental quantum state is not \emph{any} tensor network, instead it should be a highly constrained tensor network so that the emergent spacetime geometry from it should obey Einstein's equation. AdS/CFT should also emerge from the same mechanism, but in that case we need to consider the background geometry of the boundary state.

Do we have a \emph{natural} correspondence between a quantum state and a tensor network? Yes. From our former discussion, we know this must be related with quantum complexity. Piecing up all the observed mosaics, our proposal for the QM=GR mechanism is, \textbf{\emph{spacetime geometry is the optimal generative network of quantum states}}. This means\\

\textbf{\emph{The correspondence between a fundamental quantum state $\rho$ and a spatial geometry is achieved by an optimal tensor network to generate the state $\rho$. From quantum computation point of view, this optimal tensor network can be the most efficient quantum circuit to generate $\rho$ from a reference low complexity state, for example a product state for a pure fundamental state. So the tensor network is a geodesic on the quantum operation manifold w.r.t. a metric. The tensor network can build a spatial geometry as a coarse grained spin network. AdS/CFT duality can be derived from the same principle with a different complexity metric, which takes the boundary geometry into consideration. The tensor network in AdS/CFT as a geodesic can lead to an erasure QEC-like property so that the causal wedge/entanglement wedge reconstruction of bulk operations can emerge. Spacetime geometry in both QM=GR and AdS/CFT is the evolution of the tensor network w.r.t. the evolution of $\rho$, just like the spacetime geometry from spin foam. What's more, we now have two spacetime geometries. One is built from the evolving geodesic tensor network. The other is from the evolution of the fundamental quantum state, which is a quantum computation so that a spacetime geometry can be built from it following Lloyd's computational universe strategy. These two geometries should be identical. }}\\

\section{QM=GR}

\begin{figure}
  \centering
  \includegraphics[width=9cm]{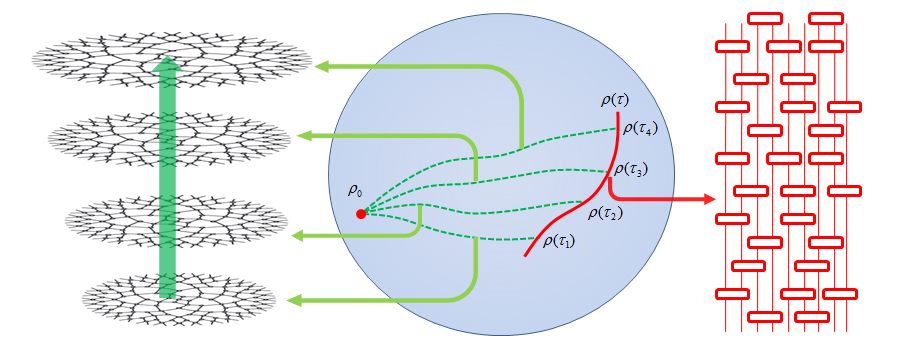}
  \caption{Spacetime as optimal generative network of quantum states. For an evolving quantum state trajectory $\rho(\tau)$, each quantum state $\rho(\tau)$ has a correspondent curve, which is generated by a geodesic quantum circuit, on the Hilbert space connecting $\rho(\tau)$ with an initial simple quantum state $\rho_0$, where the geodesic quantum circuit is defined by a metric on the quantum computational complexity of Nielsen\cite{Nielsen_geometry2}. The geodesic as a quantum circuit can be represented as a tensor network which generates a spatial geometry. The evolving quantum state $\rho(\tau)$ then corresponds to a spacetime. This picture may be closely related with spin network and spin foam in loop quantum gravity. On the other side, the evolving trajectory $\rho(\tau)$ itself as a quantum computation procedure can also generate a spacetime geometry from computational universe\cite{Lloyd2012A}.These two geometries should be identical.}\label{Fig_1}
\end{figure}

If QM=GR is correct, then we need to give solutions to the long-standing problems between QM and GR, for example

\begin{itemize}
  \item Does spacetime geometry origin from a single point(a fixed quantum state) or a curve in the fundamental Hilbert space? It's very natural to assume that an evolutional quantum system (a curve in the Hilbert space) generates a spacetime. But the problem of time in the general relativity seems to support the possibility that a single point may also lead to a spacetime. In this work, we assume that a curve in the Hilbert space determines a spacetime since a single point can be regarded as a trivial curve.
  \item Does the space geometry depend on the shape of the curve? According to the QM/GR correspondence as BEG\cite{Cao2017Bulk}, given a certain quantum state, its information pattern determines a space geometry. This is also the idea of EHM, AdS/CFT and machine learning geometry from entanglement\cite{You2017Machine}. But is this true? Is that possible that the quantum state itself can not determine the space geometry, instead the space geometry stems from the curve? If two curves in the Hilbert space intersect in one point, can this point determine a unique space geometry no matter it's reached along which curve?
  \item Can every curve correspond to a spacetime geometry? Can both pure states and mixed states have dual spacetime geometry? For an infinite quantum state, states with different quantum phases may lead to completely different spacetime?  What if a curve connects quantum states with different quantum phases in an infinite quantum system?
  \item What's the geometrical consequence of quantum properties?  For example, if two states have their correspondent geometries, then what's the geometry of their superpositioned state? How the geometry changes if a quantum tunneling happens? Some researchers believe the spacetime geometry is a sum-of-spacetime-history\cite{Lloyd2012A}. Then what's the relationship between this picture and the quantum superposition? How the superposition and evolution of quantum states will result in the geometry and matter distribution?
\end{itemize}

\subsection{$\rho\rightarrow TN(\rho)\rightarrow Geo(TN(\rho))$ from quantum state complexity}

The first component of $\rho\rightarrow TN(\rho)\rightarrow Geo(TN(\rho))$, $\rho\rightarrow TN(\rho)$, aims to map a quantum state to a tensor network, which is both a representation of the quantum state and a spatial geometry that is consistent with the information pattern of the quantum state.

The relationship between quantum complexity and quantum information pattern has been known for a long time. For example, for a 2-qubit mixed state, its Bures distance to the nearest separable state is linked with its entanglement of formation(EOF) analytically\cite{Streltsov2010Linking}.

 When quantum state complexity is concerned, usually we use the concept of quantum computation. Given a quantum state, based on the work of Nielsen to define the complexity of quantum computation algorithms $U$ as the geodesic length connecting the identity operation $I$ and $U$ for a given metric, the quantum state complexity of a quantum state $\psi$ can be defined as the minimal quantum complexity of all the quantum computations that can generate $\psi$ from a product state. The concept of quantum state complexity has been used to explain the spacetime geometry, for example complexity=volume\cite{Susskind_ER_bridge}, complexity=action\cite{Brown2015Complexity}, quantum state complexity and AdS/CFT\cite{Chapman2017Towards}\cite{Miyaji2016From}\cite{Caputa2017Anti}, quantum state complexity and time\cite{Dong_time}. Quantum state complexity is also related with quantum phase transition\cite{Chen2010Local}. So quantum state complexity should play an important role in relating quantum information concepts with spacetime geometry.

A quantum circuit usually implicates a sequence of operations, from which a concept of \emph{time} emerges. But here we are dealing with the spatial geometry corresponding to a fixed state. So we choose the tensor network picture which can be regarded as a parallel generative network of a quantum state so that time plays no role here. From the tensor network point's of view, the freedoms of the fundamental quantum state are the \emph{dangling-legs} and the bulk freedoms are the DOFs of the tensors. In the language of deep learning, the bulk freedoms correspond the structure and the parameters of the generative deep network, and the fundamental quantum state is the output of the generative network.

Due to the fact that there is no generally accepted definition of the complexity of tensor networks, here we stick to the complexity of quantum computations. That's to say,\textbf{\emph{ we define the optimal generative tensor network of a quantum state as the tensor network corresponding to the optimal quantum circuit that generates the state}}. We choose to use quantum computation complexity not only because it's well defined, we think it's more physical. For tensor networks, a state can be efficiently represented by a tensor network does not mean we can efficiently compute the state by contracting the tensor network. This is similar with the neural network beasd representation of quantum states. On the contrary, the quantum computation based computation complexity is more physical. A low computational complexity state does mean that we can generate it efficiently.

Generally speaking, the quantum state complexity of $\psi$ is defined as the minimal quantum algorithm complexity of all quantum computation algorithms that can generate $\psi$ from a \emph{simple} quantum state, where the quantum computation complexity is defined as a geodesic on a Riemannian manifold. This definition depends on two key concepts: the Riemannian metric and the definition of the initial \emph{simple} quantum states. In Niesen's work, the metric is defined so that only quantum gates with a limited size are allowed, for example only one and two-qubit gates. But a two-qubit gate can be implemented on \emph{any} pair of qubits. Following this example, generally we can define the Riemannian metric on quantum algorithm complexity by two factors: the \emph{size} and the \emph{range} of allowed \emph{elementary} quantum operation. The size means how many qubits can interact with each other and the range means which qubits are allowed to interact. In the traditional Fubini-Study metric of Hilbert space, both the size and the range are the same as the system size, which we denote as the infinite-size-infinite-range metric. With this metric the manifold of quantum computation algorithms has a finite radius since the maximal distance tween any two states are bounded by $\pi/2$. In Nielsen's definition, the size is upper-bounded by two but the range is not constrained. So under this finite-size-infinite-range (k-local) metric, quantum algorithms can have either a polynomial or an exponential complexity and accordingly the same with quantum state complexity. If we also set constraints on the range of quantum operation, for example if we only allow local operations of a quantum state in a d-dimensional Euclidean space, then both the size and range are limited. Obviously different metric will lead to different geodesics. This will result in different tensor networks and different geometries.

Another issue is what's the initial simple state of the quantum computation to reach the target quantum state. For pure states, the natural choice is a product state. In \cite{Dong_time} the quantum state complexity of a mixed state $\rho$ is defined as the geodesic distance to the 0-complexity state which has the same spectrum of $\rho$ but with a diagonal density matrix. The result of \cite{Dong_time} seems that such a definition of initial reference state for the quantum complexity of mixed state is consistent with the concept of time. Under such a choice, we now have both simple and complex pure/mixed states.

Now we have to answer, what's the complexity metric of QM=GR? Is it an infinite-size-infinite-range metric or a finite-size-infinite-range metric? If we allow to implement a quantum computation using infinite size and infinite range operators, then the allowed quantum operations are generated by any Hamiltonian on the Hilbert space so that the Hilbert space has a finite radius and any two states has a upper-bounded geodesic distance. This is to say, any pure states is trivially represented by a single tensor. Accordingly the geometry built by the correspondent geodesic with this metric has an upper-bounded size. This leads to a trivial geometry since the single tensor can be regarded as an extremely coarse grained spin network and there is no distributed bulk freedom as expected. Here all states in the Hilbert space seem to give similar geometries, at least they have the same topology. This trivial geometry does not seem to fit our observation of our universe. If we choose the finite-size-infinite range operators, then this is the case of Nielsen's definition of quantum computation complexity. Here the geometry is much richer since now we have both polynomial and non-polynomial quantum state complexity. We also have different quantum phases for infinite quantum systems so that the quantum phase transition may have a geometric picture. As proposed by Susskind, this quantum complexity can support the geometry of blackholes and wormholes. Also our former work on the relationship between time and quantum state complexity is also based on such a definition. In \cite{Chapman2017Towards} a FS metric based complexity computation of quantum field states shows that if we set a constraint on the Lie algebra of the generator of the operators, a hyperbolic emergent space appears similar to the picture of AdS/CFT and the almost negative curvature in Nielsen's work. So the finite-size-infinite-range metric seems to be the winner.

But this is not the end of the story. From the generative network point of view, \emph{any} non-optimal generative network can also generate the fundamental quantum states and a correspondent spacetime. Why do we choose to build the spacetime by the \emph{minimal} generative network of fundamental quantum states w.r.t. a metric? And why is the finite-size-infinite-range metric preferred? Of course a direct answer to the first question is that, for a given metric the optimal generative network emerges from the minimal action principle. But for the second problem, what's the optimal finite-size to generate a geometry? What's the difference if we choose to allow operations with different sizes, 2-qubit or 3-qubit operations? We think it should be 2-qubit operations. On one side, 2-qubit operations are the minimal operations to generate general quantum states. On the other side, this is related with the extraction of the spacetime information.

The generative tensor network with its DOFs $x$ will generate the fundamental quantum states $\rho$ and the spacetime geometry as well. So $x$ encodes the spacetime geometry. According to \cite{Matsueda2014Derivation}\cite{Matsueda2014Geodesic}, the second derivative of the quantum entanglement of $\rho$ will lead to the Fisher information metric and the precision of inferencing $x$ from $\rho$ is bounded by the Fisher information matrix. Choosing the smallest 2-qubit operators and the optimal geodesic generative network can both improve the information inference accuracy and therefore help to ensure the locality of the spacetime  DOFs.

Given $\rho\rightarrow TN(\rho)$, $TN(\rho)\rightarrow Geo(TN(\rho))$ connects a tensor network with a spatial geometry.  In \cite{Evenbly2011Tensor} the relationship between two geometries of ground states of local Hamiltonians on D dimensional lattice was investigated. The physical geometry is generated from the interaction pattern of the Hamiltonian and the holographic geometry is built on the tensor network to approximate the ground state of the Hamiltonian. It was claimed that when the correlation length is of the order of the lattice spacing, then the physical and holographic geometries are identical. Our work can be regarded as an extension of \cite{Evenbly2011Tensor}. Our geometry from the generative tensor network and the geometry from computational universe are just the generalizations of the holographic geometry and the physical geometry of \cite{Evenbly2011Tensor}. We extended \cite{Evenbly2011Tensor} by explicitly defining the optimal tensor network based on quantum state complexity and building the spacetime geometry instead of only the spatial geometry as in \cite{Evenbly2011Tensor}. Another point of view is to explore the relationship between spin networks and tensor networks as illustrated in \cite{Han2017Loop}. By regarding tensor networks as coarse grained spin networks, tensor networks can be connected with spatial geometry as in loop quantum gravity.

\subsection{Quantum complexity and AdS/CFT}

We now turn to AdS/CFT duality. Similar with QM=GR, the correspondence between the boundary state and the bulk state can also be understood as generated by the quantum complexity and geodesic concepts. The difference is that now the boundary state has a background geometry. This means the metric to define the quantum state complexity is different from the background free QM=GR case.

Due to the background geometry of the boundary states, the complexity metric should not be based on the finite-size-infinite-range operations. Instead it should be a metric defined on finite-size-finite-range (geometrically k-local) operations. This means that we only allow operations among neighbours in the background geometry. Recall that a geodesic is completely determined by the initial \emph{velocity} and the metric and the geodesic curve is described by differential equations. So along the geodesic, the different freedoms of a geodesic are highly constrained and redundant. For a finite-size-finite-range operations based metric, this means along the geodesic, the information of different freedoms will propagate among each other with a finite \emph{speed}. This finite information propagation velocity and information redundancy will play an important role in deducing AdS/CFT properties.

What's the difference between the geodesics with the QM=GR's finite-size-infinite-range operations and the AdS/CFT's finite-size-finite-range operations? Roughly the later geodesic is \emph{longer} than the former geodesic. But their complexities differ only polynomically since for a finite system, a QM/GR geodesic can always be achieved by a finite-size-finite-range operation with an extra polynomial cost.

Can such a metric lead to a geodesic (a quantum computation dual to a boundary quantum state) that has a QECC-like structure so that AdS space, the causal wedge and entanglement wedge reconstruction of AdS/CFT can be reproduced?

We first check the emergence of the dual AdS bulk. In \cite{Chapman2017Towards} on the manifold of Gaussian states, the quantum state complexity is defined with respect to momentum preserving quadratic generators which form $su(1, 1)$ algebras. Then the Fubini-Study metric factorizes into hyperbolic planes with minimal complexity circuits reducing to known geodesics, which shows a similarities with holographic complexity proposals. This is an evidence that the AdS bulk geometry can be generated by a geodesic with a certain Riemannian metric. Another similar work \cite{Caputa2017Anti} computed the complexity of CFT states based on  Liouville Action to optimize the tensor network representing the bulk. Their results confirmed the AdS geometry, entanglement wedge and the back-reaction mechanism of general relativity.

Now we show the geodesic may have QECC-like properties, which can be regarded as the origin of the QECC picture of AdS/CFT. Firstly we know the information along a geodesic is highly redundant since the complete geodesic is completely determined by the initial velocity and the metric. The information redundancy is the soul of QECCs. As a trivial example, the geodesic in 3D Euclidean space is a line. Obviously the x direction is completely correlated with both y and z directions. So along the line, x information can be reconstructed by either y or z information at the end point of the line. This is very similar with the erasure code picture of AdS/CFT. Similarly, if we are working with a N dimensional system, the information of a specific dimension will propagate with a finite speed to all the other dimensions. This finite speed information propagation and information redundancy can firstly result in a normal causal wedge structure (taking the information propagation velocity as the light speed) and the causal wedge bulk reconstruction. Secondly, it can also result in an entanglement wedge reconstruction just as achieved by a QECC. So the QECC picture of AdS/CFT emerges naturally as a property of the quantum computations determined by geodesics, which results in the finite correlation propagation and RT formula, the AdS bulk geometry, the separation of UV and IR freedoms, the causal wedge and the entanglement wedge. So we also built a holographic representation of all the boundary states. But it need to be noted that a certain generative network structure can only generate a subspace of the boundary Hilbert space. This is similar to the concept of the code space of QEC. But here we are not restricted to a QEC, instead a more general generative model which shows a QEC-like property. What's more interesting, a geodesic is \emph{bidirectional}, which means along the radial direction there is no information loss and the QEC-like property holds in both directions. This is exactly what the perfect tensor network, bidirectional holographic code (BHC)\cite{Yang2016Bidirectional} and the random tensor network (RTN)\cite{Hayden2016Holographic} aim to achieve. A detailed comparison with BHC and RTN will be given in the discussion session.

In Swingle's AdS/MERA duality, MERA is regarded as a discrete version of the AdS bulk. Is MERA a quantum computation correspondent to the geodesic of the boundary state? Since MERA is only valid for a special subspace of the boundary Hilbert space, it's possible that the structure of MERA is a discrete geodesic for these boundary states. Or if we stick to Nielsen's quantum computation picture, the geodesic itself is continuous, and cMERA can be really just a continuous geodesic which connects an initial product state to the boundary state by a continuous state evolution. In \cite{Matsueda2014Geodesic} it's verified that the geodesic distance in Fisher information space is consistent with the RT formula in $CFT_{1+1}$.

Another observation related to this point is from a totally different research field, the template matching problem in computational anatomy\cite{Bruveris2011The}\cite{Bruveris2015Geometry}. There the goal is to match two images by a diffeomorphic transformations. This problem can also be formalized as to find a geodesic (with a metric on the deformation energy which can be understood as the length of a deformation curve) on the manifold of diffeomorphic transformations $Diff(\Omega)$ that connects an initial identity transformation and a target transformation. Using this geodesic on $Diff(\Omega)$, one image can be smoothly deformed to match the other image. We can immediately see this is exactly the same problem as Nielsen's quantum computation complexity theory. If MERA is a geodesic and we admit the similarity between the diffeomorphic image matching and quantum state complexity, then the optimal diffeomorphic transformation trajectory should be a coarse-to-fine deformation procedure. This means it will first compensate the global shape difference followed by fine-tuning using local deformations, which is the analogy of the structure of MERA. We have to admit that this is not a typical deformation pattern in pure diffeomorphic matching algorithms such as LDDMM and geodesic shooting\cite{Beg2004Computing}\cite{Miller2006Geodesic}. But we do not take this as an evidence to claim that MERA is not a geodesic since we do have different metrics in these two problems. It can be observed that in the metric of template matching problem, a large scale deformation is assigned by a heavy cost since it has to be implemented on each pixel of the image. A local deformation in the early phase of the geodesic has a very limited influence on the later deformation since we have friction there. But in quantum state complexity, an allowed operator always has the same cost no matter where it's located along the geodesic. But an \emph{early} local operator can propagate rapidly along the geodesic so it works as a global operator at the end of the geodesic. Under such a complexity metric, it's reasonable that a geodesic show a coarse-to-fine MERA structure.

\subsection{Quantum complexity and gravity}
In both QM/GR and AdS/CFT, a linear gravity equation, which describes the interaction between geometry and energy, is shown to be consistent with the quantum information picture. In BEG, it's proven that an emergent entanglement equilibrium can be derived from Einstein's equation (and RT formula, entropy first law) so that they are consistent. In our QM=GR hypothesis, the spacetiem geometry can be built from a quantum computation in a similar way with Lloyd's computational universe. But it need to be pointed out that computational universe is based on Einstein's equation so the gravity is not emergent there. We prefer to see an emergent gravity so that the law to describe the interaction between energy and geometry can be derived from quantum information.

In fact the geodesic mechanism may fulfill this requirement. According to Susskind's complexity=action, quantum complexity and geodesics are closely related with actions. While gravity equation can be derived from variational rules from actions. So intuitively, geodesic is an extreme of quantum complexity (action), so is gravity equation. If energy (matter) corresponds to the IR freedoms and geometry is given by the UV freedoms, the interaction between energy and geometry is just the interaction between the initial condition (the starting point) and the geodesic curve in a perturbation case. So if we fix the fundamental quantum state, a small perturbation on the initial point (the matter) will lead to a change of the geodesic(geometry). We believe this picture may also derive the entanglement equilibrium. The entanglement equilibrium gives an interaction between the IR and UV entropy under a volume-invariant perturbation. In the language of complexity=volume, volume invariance is just complexity invariance, or the geodesic length invariance. The entanglement equilibrium is consistent with gravitation equation, so they should have the same origin.

In the above deduction chain, the key component is the duality between complexity and action. This assumption itself is actually to choose the optimal/geodesic generative network. Only with this condition, variation of complexity can lead to Einstein equation as in \cite{Czech2018Einstein}\cite{Ge2018Quantum}\cite{Matsueda2014Derivation}. In another word, complexity is an intermediate agent linking QM and GR. We believe it's desirable to directly derive gravity from quantum states as in \cite{} where the gravity emerges directly from the second derivative of quantum entanglement of quantum states.

In fact, if the optimal tensor network is a geodesic, the back-reaction of general relativity is just a natural property of a geodesic. A simple picture for this point is, a local bulk excitation will generate a new boundary state, so the geodesic to reach this boundary state has to be adjusted to find the correspondent new geodesic just as explained in \cite{Miyaji2016From}.

As a conclusion, spacetime emerges from a curve in the fundamental Hilbert space as the geodesics to the points along the curve. It's the variational property of the geodesics that determines the entanglement equilibrium and the gravitation equation. This spacetime structure has some similarity with the spin foam based geometry. It's an interesting task to find their connection.

\subsection{Quantum complexity and spacetime geometry}
According to our hypothesis, for a given fundamental state, quantum state complexity based mechanism can find the correspondent tensor network as its optimal generative model. The bulk spatial geometry is then given by this tensor network. The recent work revealing the connection between tensor networks, spin networks and geometry\cite{Han2017Loop}\cite{Evenbly2011Tensor} shows the feasibility of this mechanism. Intuitively the spacetime geometry is just an evolving tensor network, which corresponds to an evolving fundamental quantum state. From the spin network picture, this evolving tensor network just corresponds to the spin foam model of spacetime. But now the spin foam has a picture of a collection of geodesics corresponding to a curve of quantum states. What's more, the path integral picture of spin foam is exactly the same as the path integral picture of the fundamental quantum state. That's to say, any path in the quantum state space has a correspondent path in the geodesic tensor network space.

\subsection{Quantum complexity and deep network}

 In \cite{Dong_deep} we have pointed out the quantum state complexity and deep network have the same geometric structure. In our hypothesis, the bulk geometry is regarded as a generative deep network for the fundamental quantum state. Now we show between the deep generative network and our generative network picture of QM=GR, we have a dictionary.

 Still we take a generative deep network of natural images in the deep learning based artificial intelligence as an analogue of this.

 A generative deep network with a certain network structure can generate different images from a fixed random input vector by adjusting its network parameters, for example the kernel patterns in a deep convolutional neural network\cite{Ulyanov2017Deep}. For a certain image, there exists an optimal deep network that can generate it most efficiently. We immediately see this is exactly the same story as in the QM=GR problem. So we have the following dictionary. \\

 \begin{tabular}{|c|c|}
   \hline
Deep generative network &  QM=GR \\
\hline
An image &  A quantum state \\
Generative networks & Quantum circuits \\
Initial input vector & Initial state \\
Optimal generative network  & Optimal quantum circuit \\
   \hline
 \end{tabular}\\

 What's more, from the output image of a network with a fixed structure, we can inference the network parameters. This parameter inference capability is given by the Cramer-Rao bound, with the Fisher information matrix playing a key role. Roughly Fisher information matrix describes how sensitive the output image changes w.r.t the change of the generative network parameters. Inferencing the network parameters from output images is the same as the reconstruction of the bulk operators from boundary states. The parameter inference accuracy determined by the Fisher information matrix might correspond the fundamentally limited spatial resolution, i.e. the Planck length. We believe taking this analogy can help us to better understand QM=GR since we have quite some understanding about deep networks.

\subsection{Geometry from tensor networks and geometry from computational universe}
Now we check the spacetime generated by an evolution of a quantum state. In the GR side, we see an evolutional optimal tensor network generates a spacetime. On the QM side, the evolution as a quantum computation can also lead to a spacetime geometry as a computational universe strategy\cite{}. What's their relationship? Intuitively they should be identical. Are they? A qualitative check is like this. The tensor network based geometry emerges as a geodesic w.r.t. quantum state complexity. Quantum complexity can lead to gravitation equation. The computational universe builds the spacetime geometry based on gravitation equation. So they essentially follow the same rule. Of course this is not a proof, but we see at least there should be no inconsistency between the two geometries. So we have two pictures on building spacetime geometry and they are also just two sides of the same coin. In computational universe, matter field can be easily derived. This may help to understand how matter emerges in the tensor network based picture.

\section{Discussion}

Based on the proposed QM=GR mechanism, we can try to address some interesting problems.

\subsection{Has every quantum state a correspondent geometry?}
In BEG and AdS/CFT, we only consider special quantum states, for example states obeying area law or the redundancy constraints. They can be regarded as simple quantum states which fall in the \emph{physical corner of Hilbert space}. Roughly speaking, they are states that can be generated with a low computational complexity. But according to QM=GR, there is no evidence to exclude the possibility that more \emph{general} states also have correspondent geometry. For a \emph{general} complex quantum state, which does not obey the area law or the redundancy constraint, it also has its optimal generative model and therefore has a geometry. If this is the case, what's the geometry of a state with a non-polynomial complexity? What's the geometry of a mixed state? What's the difference between two states in different quantum phases, which means that they can not be transformed between each other by an efficient quantum circuit according to \cite{Chen2010Local}?

For any fundamental pure state, there always exists a geodesic that leads to a spatial geometry. The only problem is that the area law for simple states does not hold any more. For example in AdS/CFT, for a high complexity boundary state, the entanglement entropy between subsystem $A$ and its complementary system $\bar{A}$ is still proportional to the extreme bulk surface separating $A$ and $\bar{A}$, but the area of the extreme surface is not constrained by the boundary area between $A$ and $\bar{A}$.

From the same picture, the geometry of the superpositions of fundamental quantum states $\rho_1$ and $\rho_2$ can be totally unrelated with the geometries of $\rho_1$ and $\rho_2$. In fact the quantum complexity is nonlinear, therefore the correspondence between QM and GR is definitely nonlinear.

Another issue is the concept of wormwholes. A wormhole seems to be an extrordinary geometric structure that can violate normal spacetime geometry of the bulk. In ER=EPR, eternal black holes are dual to a thermal field double state in the CFT side, which is essentially an entangled state. But if the boundary state is just the thermal field double state and its bulk geometry is an ER bridge, then the wormhole breaks the structure of whom? Should we still call such a bulk geometry a wormhole? Maybe the word wormhole should be used to note the correspondent geometric perturbation induced by entanglement between bulk freedoms on a \emph{normal} bulk geometry since the entanglement between bulk freedoms will violate the locality of the bulk system, which can be regarded as the existence of a wormhole.

For a mixed fundamental state $\rho=U\Lambda U^{+}$ with a spectrum $\Lambda=[\lambda_1,\lambda_2,....,\lambda_N]$, its quantum state complexity can be defined as the quantum complexity of the operator $U$. This means the initial state of the geodesic is the mixed state with a diagonal density matrix which is the same as the spectrum of $\rho$. This definition has been used in \cite{Dong_time} to show the relationship between quantum state complexity and the thermodynamic arrow of time. With this definition, the RG operation in AdS/CFT will result in a diagonal density matrix. This is an extension of the completely mixed state at the coarse grained sites, which corresponds to something like a black hole\cite{Swingle2012Constructing}. From the complexity point of view, a general mixed state with a diagonal density matrix is highly complex and there is no efficient algorithm to generate it. So the geometry of the mixed state $\rho$ is a normal geometry (generated by $U$) and a blackhole-like geometry in the center. In the Fisher information picture, the state $\Lambda$ can be regarded as a colored Gaussian noise and a completely mixed state is a white Gaussian noise. From the information inference point of view, for a white Gaussian noise, the noise signal (the completely mixed state) does not encode any information of the complex generator network besides the size of the network. So we have the worst performance to inference the structure of the generative model from the quantum state $\Lambda$. This can be understood as that we can not read the interior information of a black hole.
For a colored Gaussian noise, this is a similar situation as a black hole. But now we may inference more information than from a black hole. What's the geometry of such a general diagonal mixed state is not clear to us.

So for a simple pure state, we may have the optimal information inference performance of its generative network, i.e. the bulk geometry. This is the Planck length. For a completely mixed state, we have the worst inference capability, which corresponds to a black hole.

%

\subsection{Least action principle or path integral?}
In our QM=GR road map, the geodesic picture is explicitly obtained by a least action principle since complexity equals action. The equivalence between the least action principle and the path integral seems to support that this is also a result of path integral, which means the optimal generative tensor network is also a result of a path integral on \emph{ALL} tensor networks(or quantum circuits) that can generate the fundamental quantum state. In the path integral picture, all the pathes are equal, which means every spacetime history is equal. But which spacetime emerges finally depends on the definition of the action, or the quantum complexity metric. This observation seems to give a more concrete evidence that spacetime as the complexity based optimal generative tensor network is natural.

\subsection{Comparison with bidirection holographic code}
BHC tries to build a bidirectional correspondence between the boundary and the bulk system using pluperfect tensors. But it has problems to deal with local bulk operators, where a nontrivial local bulk operator can only be defined with a fixed background tensor network. In BHC it also mentioned, that a boundary Hamiltonian can be mapped to a  bulk Hamiltonian but the bulk locality can not be kept. From our geodesic tensor network picture, this is very natural since the local bulk freedom in BHC can be regarded as a subsystem of the initial point of the geodesic and it's coupled with the structure of the tensor network. This is to say, we can not locally change the initial point of a geodesic while keeping the shape of the geodesic fixed. In fact this is just the interaction between matter and geometry of general relativity.

Another issue discussed in BHC is the gauge invariance. The key idea is that the gauge invariance origins from the different equivalent representations of a boundary state. Then what's the Lorentz invariance and diffeomorphism invariance in our geodesic picture?

It's well known that Lorentz group is closely related with spin representation. We speculate that the Lorentz invariance stems from the equivalent representation of quantum states under local qubit operations. Lorentz group in the $(1/2,1/2)$ spin representation is generated by local operations on 2 qubits with generators given by $IX-XI, IY+YI,IZ,i(IX+XI),i(IY-YI),i(IZ+ZI)$ as the generators for boost and rotations respectively. They are essentially single qubit operations. According to the work of Nielsen\cite{}, single qubit Hamiltonian is constant in a geodesic. So a local single qubit operation on the bulk freedom (as a subsystem of the initial point of the geodesic) can be compensated by a constant single-qubit Hamiltonian on the geodesic tensor network so that the end point of the geodesic does not change. This single qubit Hamiltonian changes the structure of the tensor work and therefore changes the geometry of spacetime. So Lorentz invariance can be understood as a duality between a local single qubit operation on the start point of the geodesic and the shape change of the geodesic. The constraint is that the end point of the geodesic is kept fixed. This is compatible with the gauge invariance idea in BHC.

About the diffeomorphic invariance, still it's due to the equivalent representations of fundamental quantum states. Here we can use the path integral picture to understand it. In the path integral picture, the optimal tensor network is a result of integrating all the equivalent tensor networks which generate the same quantum state. So all the paths are equivalent, the so called optimal tensor network pops out as a result of the selected metric. By changing the metric of the quantum state complexity, then we change the structure of the optimal tensor network and therefore change the spacetime geometry. But since the fundamental quantum state is fixed, all the different spacetime geometries under different metrics are all equivalent. We think the diffeomorphism invariance corresponds to the equivalence of quantum complexity metric.

Generally a spacetime metric matrix is given by $g=W\eta W'$, where the space of $W$ has a dimension of 16 and it posses a fibre bundle structure with the 6-dimensional Lorentz group as the fibre and a 10-dimensional base space. We have seen the 6 generators of the Lorentz group and now we need to find the 10 generators of the base space. In fact they are given by $II, XX,YY,ZZ, XZ+ZX,YZ-ZY,ZY-YX,i(ZX+XZ),i(YZ-ZY),i(XY-YX)$. An easy check show that they are the generators for scaling and boosts in the 4-dimensional spacetime beyond the Lorentz group. We can see now they correspond to real 2-qubit operations. So changing the weights of these generators in the quantum complexity metric as in \cite{Nielsen_geometry2} will result in different optimal generative tensor networks but they represent the same fundamental quantum state and therefore give the same physics.

\subsection{QM=GR and surface/state correspondence}
Surface/state duality is also a proposal to connect the fundamental quantum state and a tensor network based geometry. In surface/state duality, different bulk surfaces, including closed trivial spacelike surfaces, open surfaces or nontrivial surfaces, all have quantum state correspondence. From a tensor network point of view, this is true for closed surfaces because the tensor network enclosed by the closed surface does represent a quantum state. For an open surface it seems the correspondence uniqueness between a surface and a state is lost.  But from our generative network point of view, the tensor network is a geodesic and surface/state duality does not hold in general. In our optimal tensor network picture, a trivial closed surface is only a subspace of a segment of a geodesic, it should not have a quantum state correspondence. This is to say, the tensor network enclosed by a closed surface \emph{is} a quantum state but this \emph{correspondence} is only valid for the given geodesic so the surface/state duality is not a correspondence in the meaning of QM=GR. Because the tensor network enclosed by a closed surface is not necessarily the optimal generative network for the quantum state. We can see there is a special case where the surface/state correspondence holds as a QM=GR correspondence. This is the case where the tensor network enclosed by a closed surface is a segment of the geodesic, which connects the initial state of the geodesic and a point along the geodesic. Obviously such a closed surface has a dual QM=GR state, which is just the end point of the geodesic segment. This is also valid for a nontrivial closed surface enclosing a geodesic tensor network so that the correspondent fundamental is a mixed state. If the fundamental quantum state is a pure state, then an open surface, which is part of a closed surface that encloses a geodesic, can be regarded as a mixed state since it's just a subsystem of the dual fundamental quantum state of the closed surface. But still we can not say this is a QM=GR correspondence because this surface/state duality is only valid taking the geodesic tensor network as a background. This means the surface/state correspondence is not unique, instead it depends on the boundary state and its correspondent tensor network structure.

\subsection{The difference between GHZ and W states}
Using the proposed QM=GR mechanism, we can try to understand the difference of the geometries of the GHZ and W states.

For a n-qubit system in a GHZ state $|000...0+111...1\rangle$. In the QM=GR picture, if we define the quantum state complexity by Nielsen's metric based on the finite size and infinite range quantum operation, the quantum state complexity is n. Due to the symmetry of the quantum state, the correspondent geometry is also symmetric so that any point has the same distance to all the other points. Similarly a n-qubit W state $|1000...0+0100...0+......+0000...1\rangle$ also has the same distance pattern.

What's the difference of their geometries? Susskind showed this point by checking the wormholes constructed from GHZ and W states. He considered the GHZ and W wormholes by collapsing N copies of 3-qubits GHZ and W states into 3 blackholes, denoted as A, B and C. According to ER=EPR, the GHZ and W blackholes are connected by wormholes. For the GHZ blackhole, if we merge blackholes B and C, then we can construct a wormhole connecting A and BC, Alice and Bob jump into blackholes A and BC respectively and they can meet with each other. But if we do not merge B and C, Alice and Bob can not meet each other by jumping into blackhole A and B respectively. For the W wormhole, Alice and Bob can always meet each other by jumping into A and B(or BC). Why?

\begin{figure}
  \centering
  \includegraphics[width=7cm]{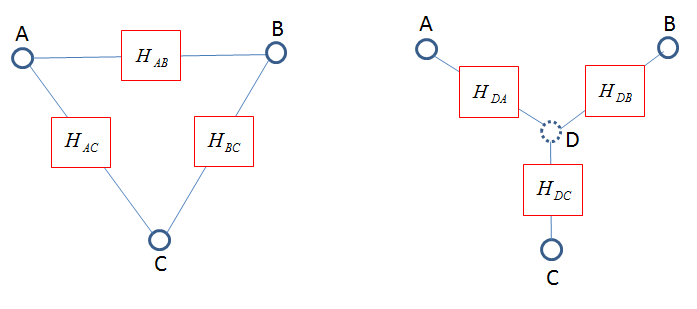}
  \caption{The optimal procedure for W and GHZ state. Left: A W state is generated by a combination of direct 2-qubit interaction; Right: A GHZ state is generated by two-qubit interaction between an ancilla qubit and all the other qubits, followed by a measurement on the ancilla qubit}\label{Fig_GHZW}
\end{figure}

We think the answer is, the geodesic quantum computations to generate a GHZ state and a W state are different. Of course the remaining problem is, can Nielsen's complexity metric can help us to prove this? We do not know how to prove this since Nielsen's quantum computation complexity is rather difficult to compute. Let alone here we really do not know the target algorithm $U$ since there exist an infinite number of $U$ that can generate W and GHZ state from a product state. But in principle, our proposed QM=GR mechanism may support such an solution. For example, a n-qubit GHZ state can be generated by preparing an ancilla qubit initialized at $|0\rangle+|1\rangle$ and carrying CNOT operations with all the n qubits and finally measureing the ancilla in the $|0\rangle+|1\rangle$,$|0\rangle-|1\rangle$ basis. Clearly this procedure is a analogue of the GHZ brane. Here all the n qubits are symmetric and there is no \emph{direct} link connecting any two of them since any two qubits can only be indirectly connected by an ancilla, which has been cut off from the GHZ brane by the measurement on it. Similarly it seems the optimal procedure to generate a W state from a product state should achieved by a combination of direct two-qubit interaction as shown in Fig. \ref{Fig_GHZW}.

\subsection{A revisit to teleportation}
Teleportation is one of the most famous quantum effect. How the quantum information can be teleported by sending two classical bit is the key issue. Here we will show how teleportation can be described in the QM=GR picture.

According to our QM=GR mechanism, spacetime geometry corresponds to certain quantum operations. Teleportation as a quantum computation procedure, of course it also generate geometries during its implementation.

The standard quantum circuit for quantum teleportation is shown in Fig. \ref{}. Alice, Bob and Tod, each one of them holds a qubit. Starting from a product state $|000\rangle$, Tod generates an unknown quantum state to be teleported. Alice and Bob entangles their qubits. After the interaction between Tod and Alice's qubits, we are at the final stage of the teleportation protocol. Before the joint measurement on Tod and Alice's qubits, the quantum state of the three qubits is generated by the quantum circuit. If we regard this circuit as the geodesic connecting the initial product state to the current quantum state, then the quantum circuit represents a space geometry. According to GM/GR, we can compute the distances between any two points on this geometry based on their mutual inforamtion. After the measurement, Tod and Alice's qubits are separated from the network shown by the red dash line in Fig. \ref{}. Now we see in the left geometry, Bob's qubit is connected with Tod's to-be-teleported qubit through a link, on which we have to interactions, $U_{AB}$ and $U_{AT}$. According to QM=GR, this link can be regarded as a kind of wormhole with a length given by $U_{AB}$ and $U_{AT}$. Among the two operations $U_{AB}$ and $U_{AT}$, $U_{AB}$ connects two points with zero distance since after $U_{AB}$, Alice and Bob's qubits are completely correlated. The real problem is $U_{AT}$. We need to reverse this operation to shrink the length of the wormhole so that Tod and Bob stand at the two ends of a wormhole with a length of 0. In fact the classical qubits from measuring Alice and Tod's qubits provide enough information to achieve this. After reversing $U_{AT}$, the length of the wormhole is zero and Bob can access the information of the teleported quantum state if Tod jumps in the wormhole. So from QM=GR, teleportation is achieved by creating a zero length wormhole connecting Tod and Bob.

Another description of teleportation is a fibre bundle or gauge picture, where the total space is the quantum states and the base space is the geometry. We regard this as a fibre bundle structure since different quantum states may correspond to the same geometry. For example, the initial and final product states of the teleportation operation essentially generate the same geometry since they have exactly the same quantum information pattern.

If geometry is determined by the geodesic tensor network, we can see permutation of qubits and local unitary operations on qubits do not change the geometry since they do not change the entanglement pattern of a quantum state. So the fibre or the gauge group is a product of permutation group and all local unitary operations on each qubit. Then obviously the operational procedure of teleportation generates a closed loop in the base space, i.e. the geometry. But in the total space we have an open curve. That's to say, teleportation achieves a kind of gauge transformation.

The initial product state corresponds to a geometry with three separated points and the gauge is fixed by the initial quantum state. After the entanglement of Alice and Bob, the geometry changes to a single point(Tod) and a link connection Alice and Bob. After the scrambling operation between Alice and Tod, all the three parts are connected. The joint measurement on Tod and Bob cuts Tod and Bob out into two separated points. At the same time, the measurement and the four possible measurement results achieve local unitary operations on the three qubit and a permutation of them as well. The post-measurement local operation on Alice inverses the four possible local operations on Bob. So finally the gauge is changed by a pure permutations of the three qubits if the post-measurement also fixes the local operations on Bob and Tod, which is of course achievable.

%
%

\begin{figure}
  \centering
  \includegraphics[width=6cm]{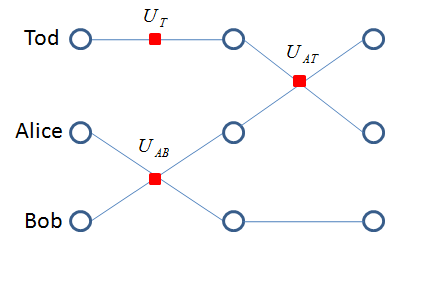}
  \caption{The geometry of quantum state teleportation}\label{fig_teleportation}
\end{figure}

%
%
The standard quantum circuit for quantum teleportation is shown in Fig. \ref{fig_teleportation}. Alice, Bob and Tod, each one of them holds a qubit. Starting from a product state $|000\rangle$, Tod generates an unknown quantum state to be teleported. Alice and Bob entangles their qubits. After the interaction between Tod and Alice's qubits, we are at the final stage of the teleportation protocol. Before the joint measurement on Tod and Alice's qubits, the quantum state of the three qubits is generated by the quantum circuit. If we regard this circuit as the geodesic connecting the initial product state to the current quantum state, then the quantum circuit represents a space geometry. According to GM/GR, we can compute the distances between any two points on this geometry based on their mutual inforamtion. After the measurement, Tod and Alice's qubits are separated from the network shown by the red dash line in Fig. \ref{}. Now we see in the left geometry, Bob's qubit is connected with Tod's to-be-teleported qubit through a link, on which we have to interactions, $U_{AB}$ and $U_{AT}$. According to QM/GR, this link can be regarded as a kind of wormhole with a length given by $U_{AB}$ and $U_{AT}$. The problem now is to shrink the length of the wormhole so that Bob can dig out the state of Tod after an operation $U_T$. Among the two operations $U_{AB}$ and $U_{AT}$, $U_{AB}$ connects two points with zero distance since after $U_{AB}$, Alice and Bob's qubits are completely correlated. The real problem is $U_{AT}$. We need to reverse this operation to shrink the length of the wormhole. In fact the classical qubits from measuring Alice and Tod's qubits provide enough information to achieve this. After reversing $U_{AT}$, the length of the wormhole is zero and Bob can access the information of the teleported quantum state since they are the same bulk point now.
%
%
%
%
%

\section{Conclusions}
In this paper we scratch a roadmap for the QM=GR hypothesis,where the correspondence between quantum states (with evolution) and spacetime geometry is based on the concept of generative models borrowed from deep learning. In our proposed mechanism, the concept of quantum complexity plays a key role. Firstly an optimal tensor network as a geodesic derived from quantum state complexity is regarded as the generative model of a quantum state, at the same time it also generates a geometry. Evolving quantum states then lead to dynamical spacetime geometry. Secondly, due to the geodesic property of the generative tensor network, the emergent spacetime geometry is consistent with Einstein's equation. Thirdly, the geodesic also has a erasure QEC-like property so that it supports the bulk reconstruction in AdS/CFT. The holographic mapping between QM and GR is just the correspondence between a quantum state and its optimal generative model. The geodesic picture of the optimal generative model is also compatible with the path integral picture so that spacetime is built from a path integral on all the generative tensor networks of quantum states. With this framework, quantum state complexity, the optimal generative tensor network and spacetime geometry have the same origin so that CA and CV hypothesises are natural. What's more, the proposed hypothesis is valid for any quantum states but not restricted to states obeying area law or the redundancy constraint. With this hypothesis, spacetime has a quantum mechanical picture so that a bulk quantum system in a background spacetime geometry can be understood as a pure quantum mechanical system. we hope this may help to enhance our understanding of the behaviour of quantum system.

 This work shows the idea of generative models for physical systems is an universal rule from microscopic quantum systems to microscopic classical world and spacetime. Maybe we should not say that spacetime is built from quantum entanglement, instead, spacetime is the optimal way to build entanglement.

\bibliographystyle{unsrt}

\bibliography{geometrycomplexity}





\end{document}